\documentstyle[12pt,aaspp4]{article}

\def\kms{\ifmmode {\, \rm km \, s^{-1}}\else {$\, \rm km \, s^{-1}$}\fi }
\def\rsun{\ifmmode {\rm R_{\odot}}\else $\rm R_{\odot}$\fi}
\def\msun{\ifmmode {\rm M_{\odot}}\else $\rm M_{\odot}$\fi}

\def\linebreak{\hfil\break}

\def\\{\hfil\linebreak}
\def\
{\hfil\linebreak}
\def\singlespace{\baselineskip=14pt}

\def\singlespace{%
    \lineskip                .15ex
    \baselineskip            3.0ex
   \lineskiplimit              0ex
   \parskip                0.60ex plus .30ex minus .15ex
   }%

\def\degpoint{\mbox{$\degree\mskip-7.6mu.\,$}}
\def\etal{{\it et al}. }
\def\msun{\ifmmode {\rm M_{\odot}}\else $\rm M_{\odot}$\fi}
\def\MSUN{\ifmmode {\rm M_{\odot}}\else $\rm M_{\odot}$\fi}
\def\msunyr{\ifmmode {\rm M_{\odot}\,yr^{-1}}\else $\rm 
M_{\odot}\,yr^{-1}$\fi}
\def\mdot{\ifmmode {\dot{M}}\else $\dot{M}$\fi}
\def\degree{\ifmmode {^\circ}\else {$^\circ$}\fi}
\def\mum{\ifmmode {\rm \mu {\rm m}}\else $\rm \mu {\rm m}$\fi}
\def\arcsec{\ifmmode ^{\prime \prime}\else $^{\prime \prime}$\fi}

\def\inch{\ifmmode ^{\prime \prime}\else $^{\prime \prime}$\fi}
\def\arcmin{\ifmmode ^{\prime}\else $^{\prime}$\fi}

\def\msun{\ifmmode {\rm M_{\odot}}\else $\rm M_{\odot}$\fi}
\def\lsun{\ifmmode {\rm L_{\odot}}\else $\rm L_{\odot}$\fi}
\def\mearth{\ifmmode {\rm M_{+\mskip-14.6muO\,}}\else $\rm 
M_{+\mskip-14.6muO\,}$\fi}
\def\mearth{\ifmmode {\rm M_{\earth}}\else $\rm M_{\earth}$\fi}
\newbox\grsign \setbox\grsign=\hbox{$>$} \newdimen\grdimen 
\grdimen=\ht\grsign
\newbox\simlessbox \newbox\simgreatbox
\setbox\simgreatbox=\hbox{\raise.5ex\hbox{$>$}\llap
     {\lower.5ex\hbox{$\sim$}}}\ht1=\grdimen\dp1=0pt
\setbox\simlessbox=\hbox{\raise.5ex\hbox{$<$}\llap
     {\lower.5ex\hbox{$\sim$}}}\ht2=\grdimen\dp2=0pt

\begin{document}

%\pagestyle{empty}

%\submitted{The Astrophysical Journal, submitted}
\title{An Optical Study of BG Geminorum: An Ellipsoidal Binary with an
Unseen Primary Star}
\author{Priscilla Benson\footnote{Wellesley College, Whitin Observatory,
106 
Central Street, Wellesley, MA 02181-8286}}
\author{Allyn Dullighan\footnote{Department of Physics and Astronomy,
Swarthmore College, 500 College Avenue, Swarthmore, PA 19081} }
\author{Alceste Bonanos$^1$}
\author{K. K. McLeod$^1$}
\author{and}
\author{Scott J. Kenyon\footnote{Harvard-Smithsonian Center for 
Astrophysics, 60 Garden Street, Cambridge, MA 02138}}
%\submitted{Received: 7 July 1999, Accepted: }
%
%\centerline{submitted to}
%\centerline{{\it The Astrophysical Journal Letters}}
%\centerline{March 1999}

%\singlespace

\begin{abstract}

We describe optical photometric and spectroscopic observations of
the bright variable BG Geminorum.  Optical photometry shows a 
pronounced ellipsoidal variation of the K0 I secondary, with 
amplitudes of $\sim$ 0.5 mag at VR$_{\rm C}$I$_{\rm C}$ and a 
period of 91.645 days.  A deep primary eclipse is visible for
$\lambda \lesssim$ 4400 \AA; a shallower secondary eclipse is 
present at longer wavelengths.  Eclipse timings and the radial
velocity curve of the K0 secondary star indicate an interacting binary
where a lobe-filling secondary, $M_2 \sim 0.5 ~ \msun$, transfers 
material into a extended disk around a massive primary,
$M_1 \sim 4.5 ~ \msun$. The primary star is either an early B-type
star or a black hole.  If it did contain a black hole, BG Gem would be
the longest period black hole binary known by a factor of 10, as well
as the only eclipsing black hole binary system.

\end{abstract}

\subjectheadings{binaries: eclipsing -- binaries: spectroscopic --
stars: emission-line -- stars: evolution -- stars: individual (BG Gem)}

\section{INTRODUCTION}

BG Geminorum was discovered by Hoffmeister (1933) and
Jensch (1938) as a possible RV Tau star with an uncertain
period of $\sim$ 60 days.  With a photographic magnitude of
$\sim$ 14, the star languished in the General Catalog of Variable
Stars (\cite{kho85}) until 1992, when we 
began photometric observations to improve the period estimate and 
to verify the RV Tau classification.  Early data revealed 
a repeatable ellipsoidal variation with a long period, 
instead of the more irregular variation expected from an
RV Tau star.  Additional photometry allowed us to refine the 
period and plan spectroscopic observations.  Together with the 
photometry, high quality spectra acquired around the orbit 
establish BG Gem as a rare eclipsing binary system with an unseen 
primary star and a lobe-filling K supergiant secondary star.

This paper describes our data and analysis.  We begin with a
summary of the observations in \S2, continue with a detailed 
analysis in \S3, and conclude with the discussion in \S4.

\section{OBSERVATIONS}

Various student observers acquired optical photometry of BG Gem
with standard VR$_{\rm C}$I$_{\rm C}$ filters and a Photometrics 
PM512 camera mounted on the Wellesley College 0.6-m Sawyer telescope.  
Observations beginning 23 July 1997 used a TK1024 back-illuminated CCD.
We processed the data using standard tasks within NOAO 
IRAF\footnote{IRAF is distributed by the National Optical Astronomy 
Observatories, which is operated by the Association of Universities 
for Research in Astronomy, Inc., under contract to the National 
Science Foundation}.  Each image was bias-subtracted, dark-subtracted,
and flat-fielded
using twilight flats.  We extracted photometry of BG Gem and several
comparison stars using PHOT with a 3\arcsec~aperture and a 
3\arcsec~sky annulus 17\arcsec~from the star.  
Wellesley is not a photometric site; we thus derived photometry
of BG Gem relative to comparisons verified to be non-variable.
The relative photometry has 1$\sigma$ probable errors of 
$\pm$0.011 mag at V, $\pm$0.014 mag at R$_{\rm C}$, and
$\pm$0.011 mag at I$_{\rm C}$ based on repeat observations of 
comparison stars in the field of BG Gem.
Table 1 lists the relative photometry of BG Gem and a
comparison star as a function of time 
(the Heliocentric Julian Date, JD) and photometric phase $\phi$ 
defined below.

P. Berlind, M. Calkins, and several other observers acquired
low resolution optical spectra of BG Gem with FAST, a high 
throughput, slit spectrograph mounted at the Fred L. Whipple 
Observatory 1.5-m telescope on Mount Hopkins, Arizona (\cite{fab98}).
We used a 300 g mm$^{-1}$ grating  blazed at 4750 \AA, 
a 3\arcsec~slit, and a thinned Loral 
512 $\times$ 2688 CCD.  These spectra cover 3800--7500 \AA~at 
a resolution of $\sim$ 6 \AA.  
We flux- and wavelength-calibrated the spectra in IRAF. 
After trimming the CCD frames at each end of the slit, 
we corrected for the bias level, flat-fielded each frame, 
applied an illumination correction, and derived a full wavelength 
solution from calibration lamps acquired immediately after each 
exposure.  The wavelength solution for each frame has a probable 
error of $\pm$0.5--1.0 \AA.  We extracted object and sky spectra 
using the optimal extraction algorithm within APEXTRACT.  The
absolute flux-calibration for each night relies on observations 
of 2-5 standard stars (Barnes \& Hayes 1982) and has an uncertainty
of $\pm$5\%--10\%.  Nights with clouds or poor seeing have poor 
absolute calibrations ($\pm$ 20\% or worse) but good relative
calibrations.

Figure 1 shows a typical spectrum of BG Gem.  The spectrum resembles
a late G or early K star, with a red continuum and strong absorption
from the G band, Mg~I, Na~I, and the Ba~I blend at $\lambda$6495. 
Strong H~I Balmer emission lines are also visible, with H10 and H11 
present on some spectra.  The H$\alpha$ and H$\beta$ lines
are often double peaked.  Some spectra show a weak blue continuum
and weak He~I emission at 5876 \AA~and 6678 \AA.  There is
no evidence for absorption lines from a hot, early-type star.

\begin{figure}[tbh]
\epsscale{0.7}
\plotone{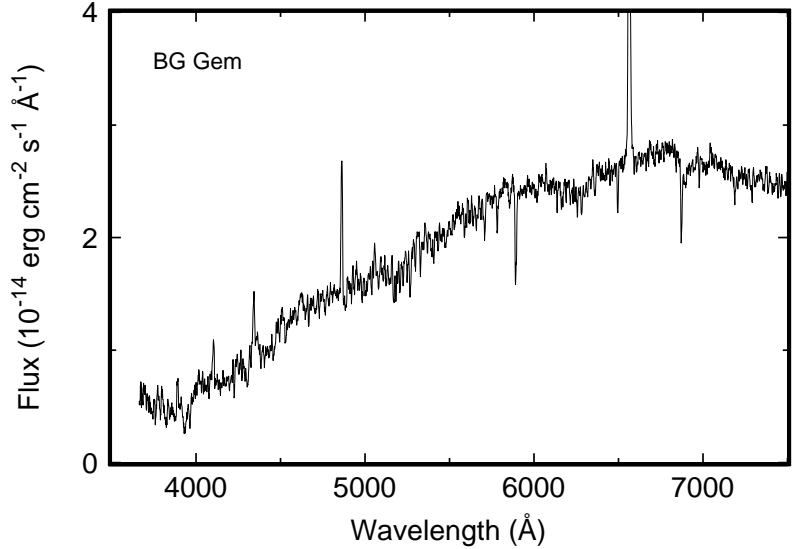}
\figcaption[Benson.fig1.eps]{Optical spectrum of BG Gem. Strong H~I emission lines
and a faint blue continuum shortwards of 4000 \AA~are produced by
the hot primary. The red continuum and prominent Mg~I and Na~I 
absorption lines are from the K0 I secondary star.
}
\end{figure}

To analyze the spectra, we measured continuum magnitudes and indices
of strong absorption and emission lines using narrow passbands
(O'Connell 1973; Worthey 1994).  Table 2 lists the central wavelength
$\lambda$ and width $\delta \lambda$ for each. The
continuum magnitudes are normalized to the zero-point of the
V magnitude scale, $m_{\lambda}$ = $-2.5~{\rm log}~ F_{\lambda}$
$-$21.1, where $F_{\lambda}$ is the flux in the passband.
The absorption and emission indices, derived using SBANDS within
IRAF, are
$I_{\lambda}$ = $-$2.5 log($F_{\lambda}$/$\bar{F}$),
where $F_{\lambda}$ is the average flux in the passband $\lambda$ and
$\bar{F}$ is the continuum flux interpolated between the fluxes in
the neighboring blue band centered at $\lambda_b$ and the red band 
centered at $\lambda_r$.
Table 3 lists the measured indices along with emission line equivalent
widths as a function of JD and $\phi$.  Table 4 lists continuum
magnitudes measured on spectrophotometric nights.

We searched various databases for previous observations of BG Gem.
BG Gem is not a known point source in any of the online {\it IRAS},
radio, or X-ray catalogs (e.g., {\it ASCA}, {\it ROSAT}).
It lies close to several triggers in the CGRO/BATSE 4B catalog
(e.g., trigger \#4157), but large positional offsets from the nominal
trigger positions makes association with these events unlikely.  
The average of 100 days of data from the All-Sky Monitor (ASM) of
the {\it Rossi X-ray Timing Explorer} is $-0.1 \pm 1.0$ mCrab at
2--12 keV (Remillard 1999; 1 mCrab is 0.0755 counts s$^{-1}$ with
the ASM (Levine \etal 1996), which corresponds roughly to
$\sim 10^{-11}$ erg cm$^{-2}$ s$^{-1}$ at 2--12 keV assuming 
a spectrum similar to that other X-ray binaries [see \cite{so99b}]).  
BG Gem is a 2MASS point source, \#0603308+274150 in the 1999 
Spring Release Point Source Catalog, as summarized below.

\section{ANALYSIS}

\subsection{The Light Curves}

Figure 2 shows light curves for BG Gem.  The data have been folded 
on the best period derived from Fourier analysis and periodograms.
The primary minimum has the ephemeris

\begin{equation}
\rm Min = JD~2449179.7437 \pm 1.037 ~ + ~ (91.645 \pm 0.315) \cdot E ~ .
\end{equation}

\noindent
We define $\phi$ = 0.0 at primary minima.
The curves at 4050 \AA~and 4400 \AA~are derived from FAST spectra.
The zero points for the V and R data are set from FAST
light curves at 5550~\AA~and 6500~\AA~using offsets derived 
from the Jacoby, Hunter \& Christian (1984) spectral atlas 
and published broadband magnitudes,
V = $m_{5550}$ + 0.16 and
R$_{\rm C}$ = $m_{6475} - 0.48$.
The zero point for the I$_{\rm C}$ light curve assumes the
secondary is a normal K0 I star reddened by $A_V \approx$ 1.65 mag
(see below), with an intrinsic R$_{\rm C}$--I$_{\rm C}$ = 0.52 
(Bessell 1990).

\begin{figure}[tbh]
\epsscale{0.7}
\plotone{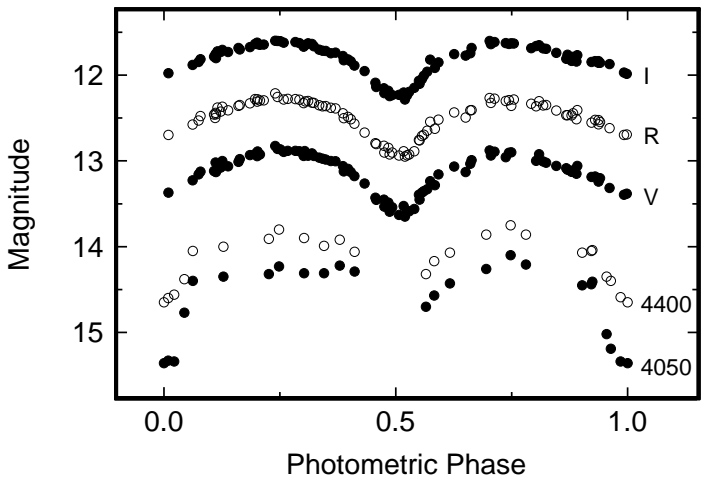}
\figcaption[Benson.fig2.eps]{Optical light curves of BG Gem.  Narrow-band continuum 
magnitudes at 4050 \AA~and 4400 \AA~are from FAST spectra.
Broadband magnitudes are from Wellesley CCD data scaled as
described in the text.  All light curves show ellipsoidal 
variations of the lobe-filling secondary and an eclipse of
the hotter primary at phase 0.}
\end{figure}

The light curves show several clear features.  Ellipsoidal variations
of the secondary star are prominent at VR$_{\rm C}$I$_{\rm C}$.  
The large amplitudes of $\sim$ 0.5 mag indicate that the
secondary probably fills its Roche lobe and that the orbital inclination 
is close to 90\degree.  
The deep primary minimum at short wavelengths shows that 
the primary star is much hotter than 
the secondary.  The eclipse of the primary is long and 
flat-bottomed, which implies that the primary is extended.
This extended primary is probably responsible for the eclipse 
of the secondary star visible in the V, R$_{\rm C}$ and I$_{\rm C}$
filters as a deep secondary minimum at $\phi$ = 0.5.

The absorption line indices shown in Figure 3 provide additional
support for a hot, extended, primary star in BG Gem.  Due to a 
decrease in the continuum flux, all of the lines are strongest 
during the eclipse of the primary.  The blue lines strengthen much more
than the red lines, indicating that the primary has a blue continuum.
The eclipse of the primary covers a significant fraction, $\sim10\%$,
of the orbital period.

\begin{figure}[tbh]
\epsscale{0.7}
\plotone{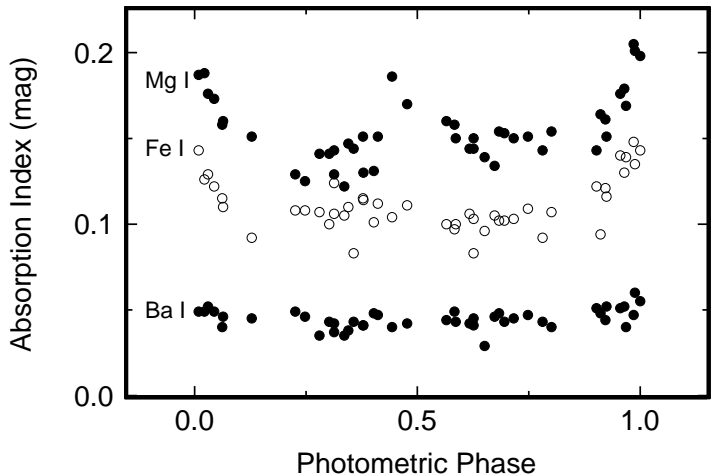}
\figcaption[Benson.fig3.eps]{Variation of absorption line indices. The Ba~I index
at 6495 \AA~is nearly constant with phase. The Mg~I and Fe~I
indices strengthen during primary eclipse.  Blue absorption
lines -- e.g., Fe~I $\lambda$ 4400 -- strengthen relatively
more than red absorption lines -- e.g., Mg~I $\lambda$ 5200.}
\end{figure}

We can measure the spectral type of the secondary star 
by examining the absorption lines at $\phi$ = 0.0, when the
primary is in eclipse.  Based on a comparison of the spectral 
indices with indices derived from the Jacoby \etal (1984) and 
Worthey (1994) spectral atlases, all of the absorption lines 
imply a late G or early K spectral type.
Our best estimate is K0 I, with an
uncertainty of 1--2 subclasses.  The supergiant classification 
is required for the combination of strong Mg~I, Na~I, and
Fe~I absorption features.  This result agrees with the low
gravity implied for a lobe-filling secondary.

The Balmer emission line equivalent widths also vary with 
phase (Figure 4).  The H$\alpha$ and H$\beta$ lines are 
eclipsed at primary minimum.  The eclipses are long and 
never reach totality.  Both lines strengthen relative to 
the continuum during secondary eclipses, because the flux 
from the K0~I secondary weakens considerably.  Similar 
variations may be visible in He~I
$\lambda$6678, but the line is weak with a typical equivalent
width of 0.5 \AA~or less.  Other He~I lines are too difficult
to measure accurately: $\lambda$5876 is blended with Na~I and
$\lambda$7065 is very weak.

\begin{figure}[tbh]
\epsscale{0.7}
\plotone{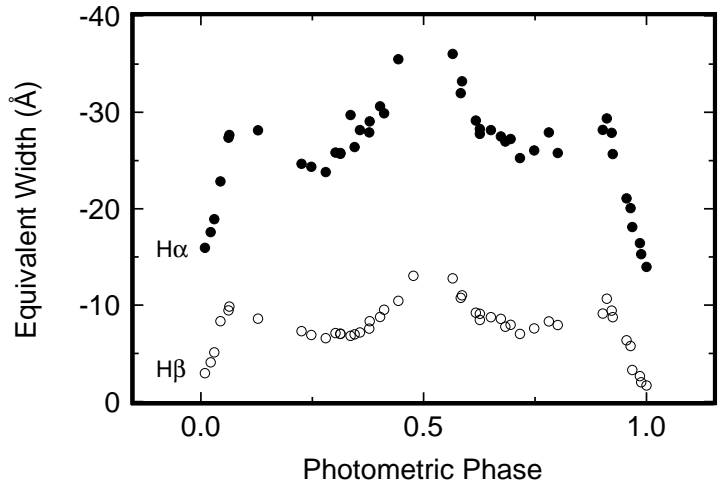}
\figcaption[Benson.fig4.eps]{Variation of emission line equivalent widths. 
The H~I emission is eclipsed by the secondary at $\phi$ = 0.
The emission flux is roughly constant outside eclipse: the 
equivalent widths rise at $\phi$ = 0.5, because the K0 
continuum declines.}
\end{figure}

The light curves constrain the sizes of the H~II emission 
region and the primary and secondary stars.  For an edge-on 
system with a circular orbit, the eclipse duration yields the 
sum of the radii 
of the two components; the duration of totality establishes 
the difference of the two radii. In BG Gem, the length of 
the eclipses at 4400 \AA~and 4050 \AA~results in
$R_1 + R_2 \approx$ 0.37 $A$, where $A$ is the orbital semi-major
axis, $R_1$ is the radius of the primary star, and $R_2$ is the 
radius of the K0 I secondary star.  From the length of totality,
$R_2 - R_1 \approx$ 0.12 $A$. These two constraints set
$R_1 \approx$ 0.12 $A$ and $R_2 \approx$ 0.25 $A$. The apparent
length of totality at $\phi$ = 0.5 in the R$_{\rm C}$ and I$_{\rm C}$ 
filters
suggests a somewhat larger radius for the primary at long
wavelengths.  The length of the H$\alpha$ eclipse 
implies $R_{\rm H \alpha} + R_2 \approx$ 0.55 $A$, which results
in $R_{\rm H \alpha} \approx$ 0.3 $A$. The lack of a total 
eclipse for H$\alpha$ confirms that the H~II region is larger 
than the K0~I secondary.

To derive absolute constraints on the binary components, 
we need an independent measure of $A$.  We now consider
radial velocity data to establish the orbital parameters
of BG Gem.

\subsection{Radial Velocities}

We derived radial velocities from the strong absorption and 
emission lines on FAST spectra.  For absorption line velocities, 
we cross-correlated the 
FAST spectra against the best-exposed spectrum, where the
velocity is set by cross-correlation against standard stars
with known velocities (see Tonry \& Davis 1979; Kurtz \& 
Mink 1998).  To avoid contamination from the hot primary,
we restricted the cross-correlation to $\lambda\lambda$5000--6800.
We measured emission line velocities from cross-correlations with
an emission-line template, as described by Kurtz \& Mink (1998).
We adopted the velocity of H$\beta$ as the emission line velocity,
because H$\alpha$ may be blended with [N~II] emission on our
low resolution spectra.
We estimate errors of $\pm$15 \kms~ for absorption lines and 
$\pm$20 \kms~for emission lines. Table 5 lists the measured 
velocities as a function of JD and $\phi$.

Figure 5 shows the absorption line radial velocities as a
function of photometric phase.  We analyzed these observations
using Monet's (1979) Fourier transform algorithm (see also
Kenyon \& Garcia 1986).  The best spectroscopic period,
$P_{spec} = 90.18 \pm 5.28$, agrees with the photometric
period.  A circular orbit with $P = P_{phot}$ from equation
(1) fits the orbit well.  This solution has an orbital 
semi-amplitude, $K_{\rm K0} = 74.6 \pm 4.5$ \kms,
and a fractional semi-major axis,
$A_{\rm K0}~{\rm sin}~i = 0.63 \pm 0.04$ AU.
Spectroscopic conjunctions occur 1.234 days prior to
photometric minima,

\begin{equation}
\rm Conj = JD~2451057.2325 \pm 0.957 + 91.645 ~ \cdot ~ E ~ .
\end{equation}

\noindent
This phase difference has a significance of only 1.3$\sigma$.
The mass function is 
$M_1^3 ~ {\rm sin}^3 i = (3.97 \pm 0.09)~\msun ~ (M_1 + M_{\rm K0})^2$.
If we adopt sin $i$ = 1, this result becomes
$M_1 \approx 4 ~ \msun ~ (1 + q)^2$,
where $q \equiv M_{\rm K0}/M_1$ is the mass ratio.

\begin{figure}[tbh]
\epsscale{0.7}
\plotone{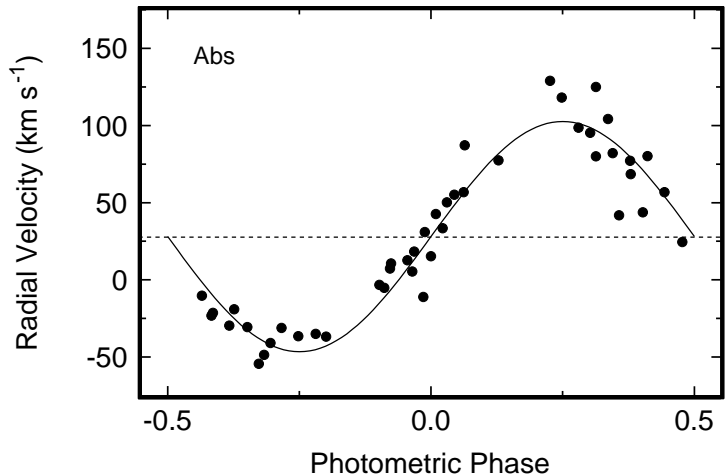}
\figcaption[Benson.fig5.eps]{Absorption line radial velocity curve.
The solid line is the best-fitting circular orbit to
the measured velocities (filled circles). The dashed
line indicates the systemic velocity.  Note that we have centered this
plot on $\phi=0.0$ for clarity.}
\end{figure}

Solutions with eccentric orbits do not improve the 
fit to the data.  Iterations in Fourier and configuration
space yield $e = 0.096 \pm 0.065$. The Lucy \& Sweeney (1971) 
test confirms that the non-zero $e$ has only 1.5$\sigma$
significance.  We thus prefer the circular solution with
the parameters quoted above.  These results place a lower
limit on the mass of the primary, $M_1 \gtrsim$ 4 \msun,
if the K0~I secondary has negligible mass and sin $i$ = 1.

The emission line velocities and double-peaked profiles indicate that the 
emission lines
form in a circumstellar disk surrounding the primary star (Figure 6).  
The H$\beta$ line has a nearly constant velocity outside primary
eclipse, with a mean velocity of +33.8 $\pm$ 16.3 \kms.  This 
velocity is identical to the systemic velocity $\gamma$ = 
28.0 $\pm$ 3.2 \kms.  The line velocity 
increases as the eclipse of the blue continuum begins.  The 
velocity reaches $\gamma + 125$ \kms~at $\phi$ = $-$0.05 to $-$0.04,
decreases to the systemic velocity at $\phi$ = 0.00, falls
to $\gamma - 125$ \kms~at $\phi$ = 0.04 to 0.05, and returns
to the systemic velocity as the eclipse of the blue continuum
ends. This classical disturbance -- the Rossiter
effect -- is consistent with the eclipse of a rapidly rotating
disk. The blue-shifted half of the disk is eclipsed first during 
ingress and is revealed first during egress.  Both H$\alpha$
and H$\beta$ decrease 
in full-width at half maximum during eclipse, supporting
this interpretation. If we associate the apparent rotational 
velocity of $\sim$ 125--150 \kms~with disk rotation,
the radius of the disk is 
$R_{\rm H \alpha} \approx$ 35--50 \rsun $(M_1 / 4 \msun)$.
Together with our limit on $R_{\rm H \alpha}$ from the
eclipse duration, this result leads to a rough estimate
for the mass ratio, $q \sim$ 0.1, and
the semi-major axis, $A \approx$ 150 \rsun.

\begin{figure}[tbh]
\epsscale{0.7}
\plotone{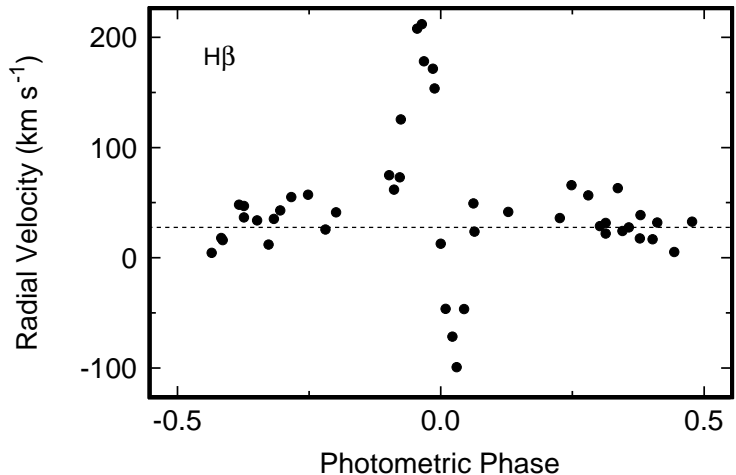}
\figcaption[Benson.fig6.eps]{Emission line radial velocity data for H$\beta$.
The dashed line indicates the systemic velocity.  Note that we have
centered this plot on $\phi=0.0$ for clarity.}
\end{figure}

The small mass ratio supports our conclusion that the K0 supergiant
fills its Roche lobe.  The effective radius of the Roche lobe
for $q = 0.1 \pm 0.05$ is $R_L \approx 0.25^{+0.06}_{-0.04} A$,
indistinguishable from our estimate for the radius of the secondary,
$R_L \approx 0.25 A$. The effective radius of the Roche lobe 
for the primary star is $R_L \approx 0.71^{+0.06}_{-0.07} A$
$\approx$ $105 \pm 10~\rsun$. The radius of the H$\alpha$ emission
region is thus roughly half of the tidal radius. The radius of
the primary at blue wavelengths 
is only $\sim$ 20\% of the tidal radius,
but large compared to the radius of a normal main sequence star.

These results suggest a simple dynamical model for BG Gem.
The system consists of a lobe-filling K0 supergiant which transfers 
material into a large luminous accretion disk surrounding 
a primary star with $M_1 \sim$ 4--5 \msun.  The disk must 
have a blue continuous spectrum to explain the deep primary 
eclipses at $\lesssim$ 4500 \AA.  This disk is also 
responsible for the secondary eclipses at longer wavelengths.  

To refine this model, we now consider the light curve and spectra
in more detail.  We begin with an analysis of the light curve
using the Wilson-Devinney (1971) code.  Improved constraints
on the orbital parameters and on the physical characteristics of 
the binary components lead to better limits on the nature of the 
accretion disk.  We then use the reddening derived from the Balmer 
lines and the optical continuum to set the distance to the system.

\subsection{Detailed Light Curve Analysis}

Wilson \& Devinney (1971) developed a modern light curve 
synthesis program to derive physical characteristics of
binary stars from multi-color light curves.  Wilson (1990,
1994, and references therein) describes recent improvements 
to the code, which is
now available by anonymous ftp.  The code consists of two
programs: LC, used to compute model light curves or radial 
velocity curves as a function of various input parameters;
and DC, which can be iterated to derive the best set of binary parameters 
from
a least squares fit of the model to actual observations.  
The analysis of \S3.1 and \S3.2 demonstrates that 
the primary component of BG Gem is an extended disk surrounding
a massive star.  We view the system at close to 90\degree.
Few light curve synthesis programs include radiation from a
disk; none reliably compute fluxes for an edge-on disk.  Therefore,
for this study,
we used LC to derive reliable ranges for binary parameters that
produce light curves similar to those in Figures 2--4, but omitted 
the iteration of these parameters with DC.  

Figure 7 shows light curves from Figure 2 along with model 
light curves. 
To determine the best fit binary system (dashed curves), we varied several parameters
to derive a `best' approximation to the observations,
(i) the mass ratio, $q$;
(ii) the inclination, $i$; and
(iii) the brightness of the primary star relative to a
lobe-filling secondary with an effective temperature,
4500 K, appropriate for a K0 supergiant star.
The mass ratio and inclination set the amplitude of the
ellipsoidal variations; the ellipsoidal amplitude grows
with sin $i$ and inversely with $q$.
The relative brightness of the primary sets the depth of 
the primary eclipse and is also constrained by the variation
of absorption line indices with wavelength.

\begin{figure}[tbh]
\epsscale{0.7}
\plotone{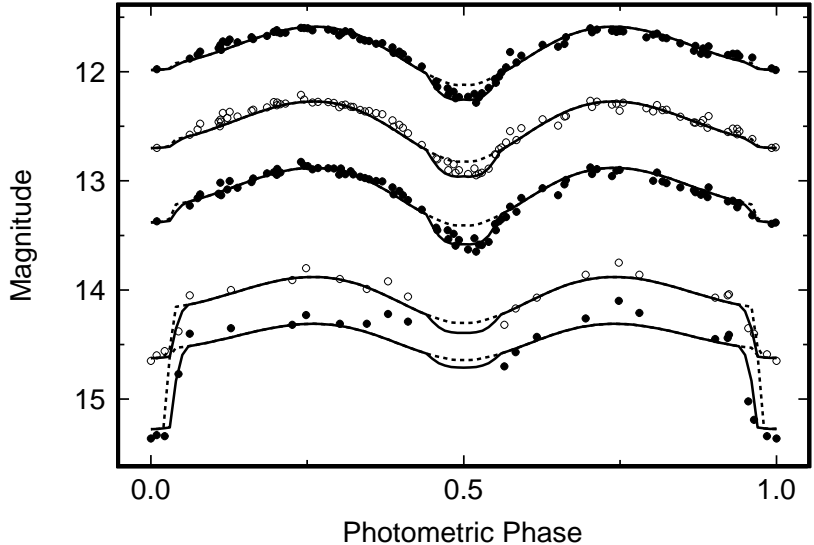}
\figcaption[Benson.fig7.eps]{Comparison of binary models with observations.
Observations are as in Figure 2.  Model light curves, dashed
lines for a binary with a small B-type primary star and 
a lobe-filling K0 star with $q = 0.1$ and $i = 90\degree$, and solid lines
for a disk-shaped primary.
At $\phi$ = 0.25, the B-type primary emits 
40\% at 4050 \AA,
25\% at 4400 \AA,
9\% at 5500 \AA,
5\% at 6500 \AA, and
3\% at 8000 \AA. Compared to the stellar primary (dashed lines)
a disk with a rectangular cross-section (solid lines)-- described in 
the text -- produces more rounded eclipses at $\phi$ = 0.
and deeper eclipses at $\phi$ = 0.5.}
\end{figure}

The dashed light curve models in Figure 7 have several successes and 
failures.  The models match the depth of primary eclipse 
at each wavelength and the overall amplitude and shape 
of the light curves at $\phi \approx$ 0.05--0.40 and 
$\phi \approx$ 0.60--0.95.  The length of the model eclipse
at $\phi$ = 0 is too short, because the model primary is a star
instead of a disk.  The secondary minimum at $\phi$ = 0.5 
is too shallow, because the small primary star cannot eclipse
very much of the lobe-filling K0 I secondary.  The primary 
star must be hot and small to match the depths of primary
eclipse at 4000--9000 \AA.

To make a more realistic light curve synthesis, we considered a crude 
disk model for the primary star (solid curves in Figure 7).  We assumed a disk with a rectangular 
cross-section as viewed in the orbital plane.  This disk has
height $H_d$ above the orbital plane and radius $R_d$ in the orbital
plane.  We assumed $R_d$ = $\beta(\phi) ~ R_L$ and $H_d$ = 0.05--0.1
$R_d$, 
where $R_L$ is the effective radius of the Roche lobe of the 
primary.  We then took the output of LC -- the fluxes of the 
primary and secondary as a function of $\phi$ -- and constructed 
new model light curves.  We computed 
(i) the fraction of the disk eclipsed by the K0 secondary star 
at $\phi = 0.0 \pm 0.1$, and
(ii) the fraction of the K0 secondary eclipsed by the disk 
at $\phi = 0.5 \pm 0.1$.  We accurately accounted for Roche
geometry and assumed that both light sources are completely
opaque and uniformly illuminated.  This approach ignored limb 
and gravity darkening of the secondary, but should yield light
curves closer to those actually observed.  

The simple disk-shaped primary can explain the depths and shape 
of both primary and secondary eclipses, as indicated by the solid
lines in Figure 7. Models with $H_d$ = 0.07 $R_d$,
$\beta(\phi = 0.0) \sim 0.2$, and $\beta(\phi = 0.5) \sim$ 0.8--0.9
yield the `best' fit to the data.  Less extended disks do not
eclipse enough of the secondary at $\phi = 0.5$.  More
extended disks occult too much of the secondary.  The large
variation in disk radius with orbital phase suggests that the
opaque, outer part of the disk is much cooler than the inner 
part of the disk.  Our model
is too simple to constrain either the size of the disk as a 
function of wavelength or the brightness temperature distribution
within the disk, as in Vrielmann, Horne, \& Hessman (1999) for
example.  We plan more detailed analyses in future studies.  

The models shown in Figure 7 assume $i = 90\degree$, $q = 0.1$, 
and a primary star with an effective temperature of 10,000 K.  
Primary stars with different effective temperatures do not 
change the model light curves as long as the ratio of the luminosity
of the primary star to the luminosity of the secondary star at each 
wavelength remains fixed.  
These ratios are set by the eclipse depth and cannot vary by 
much more than 10\%.  For fixed inclination, smaller mass ratios
produce shallower secondary mimina.  The disk radius also
shrinks as $q$ increases, which leads to a smaller ``disk 
eclipse'' at secondary minimum.  We estimate $q = 0.1 \pm 0.05$
for sin $i$ = 1.  Any geometry with $i \lesssim$ 80\degree~ does 
not produce a primary eclipse and is thus ruled out by the data.  
Improved constraints on these values using our data requires a 
WD-type code that includes a reliable model for an edge-on 
accretion disk.

\subsection{The Nature of the Primary}

To derive better limits on the nature of the primary star, we need 
good estimates for the reddening and distance in addition to the 
parameters derived above.  With $l = 183\degree$ and 
$b = 2\degpoint8$, BG Gem lies close to the galactic plane in 
the direction of the galactic anti-center.  Previous extinction 
surveys suggest modest visual extinctions, $A_V \sim$ 1--2 mag, 
for distances of 1--5 kpc (e.g., Neckel \& Klare 1980; 
Hakkila \etal 1997).   Adopting $A_V$ = 1.5 $\pm$ 0.5 mag
yields $V_0$ = 11.5 $\pm$ 0.5 for the system at maximum
and $d = 2.5 \pm 0.5$ kpc for a K0 I star with
$M_{bol}$ = $-$1.9 from the LC program and
a bolometric correction of 1.5 mag.

We can place better limits on the extinction from 
near-IR photometry and the emission line data.
A K0 supergiant should have V--J = 1.5--1.6 and
V--K = 2.1--2.3.  Data from 2MASS -- 
J = 10.32 $\pm$ 0.03, H = 9.61 $\pm$ 0.03, and
K = 9.34 $\pm$ 0.03 -- and V = 13.0 at maximum imply
$A_V = 1.65 \pm 0.25$ for a normal extinction law
(e.g., Rieke \& Lebofsky 1985; Bessell \& Brett 1988).
The mean intensity ratio for H$\alpha$ and H$\beta$
is $I$(H$\alpha$)/$I$(H$\beta$) =  5.81 $\pm$ 1.01;
this ratio suggests $A_V = 1.6 \pm 0.5$ for a standard
extinction law (Mathis 1990).
We thus favor $A_V = 1.65 \pm 0.25$ for the extinction
and $d = 2.25 \pm 0.25$ kpc.  This combination agrees 
with the distance-dependent extinction measurements 
summarized in Hakkila \etal (1997).

Figure 8 shows the observed spectral energy distribution 
for BG Gem.  Filled points indicate observations dereddened
by $A_V = 1.65$.  The K0 supergiant produces most of the
optical and near-infrared radiation, as indicated by the
solid and dot-dashed curves.  The primary source contributes a
modest fraction of the V light. Its fraction of the total 
light grows substantially towards shorter wavelengths,
as shown by the dashed curve.  The slope of the hot
continuum, $\alpha$ = (d log $F_{\lambda}$/d log $\lambda$) = 
$-2.5 \pm 0.75$, is best-fit by the spectrum of an 
accretion disk but is marginally consistent with the 
spectrum of an early type star.  The measured extent
of the primary, $R_1 \approx 18\pm2 ~ \rsun$, is much
larger than a normal early-type main sequence star.
The predicted V magnitude of a spherical star with
$T \ge 10^4$ K and $R_1 \approx 18~\rsun$ is roughly
2--3 mag {\it brighter} than the K0 secondary.  Our
observed limit on the V brightness of the primary,
$\sim$ 10\% of the K0 secondary, demonstrates that 
the primary is not a star.  However, a flattened disk with 
$H_d/R \sim$ 0.05--0.10 satisfies the limits on $R_1$ 
and the total brightness derived from the light curves. 

\begin{figure}[tbh]
\epsscale{0.7}
\plotone{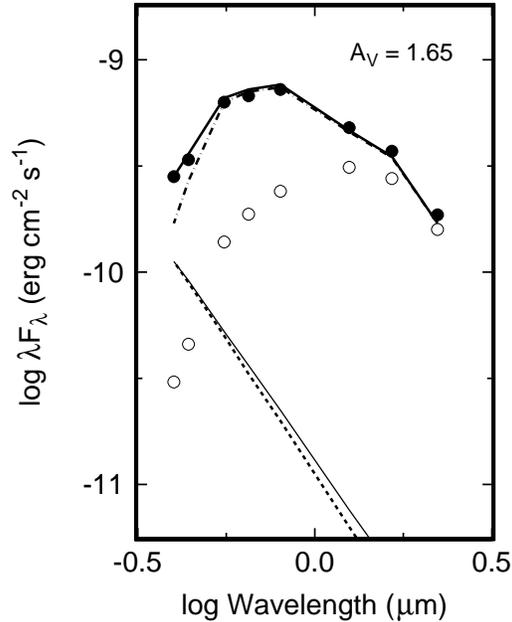}
\figcaption[Benson.fig8.eps]{Spectral energy distribution of BG Gem.
The open circles indicate observed fluxes; filled
circles indicate dereddened fluxes. The dot-dashed
line plots the spectrum of a K0 I star normalized
to emit 60\% of the flux at 4050\AA.  The dashed
line plots the spectrum of the primary,
log $\lambda F_{\lambda} \propto \lambda^{-2.5}$,
normalized to emit 40\% of the 4050\AA~flux.
The heavy solid line shows the combined flux of the
K0 supergiant and the primary. The light solid line 
indicates the spectrum of the disk model described 
in the text.}
\end{figure}

Constructing a `best' model for the disk in BG Gem requires 
ultraviolet or X-ray data to provide better constraints 
on the spectrum and on the radius of the primary star. In 
this paper, we simply develop a model consistent with the observed
spectrum and masses without considering whether it is `best' 
or unique.  We first examine the case where the disk surrounds a hot
primary 
star with parameters defined in \S3.3.  The primary
star either has a low luminosity compared to the disk 
or is occulted by the disk. Otherwise, our model cannot
account for the primary eclipse as outlined above. Requiring 
a self-luminous disk to produce all of the optical flux 
yields a large accretion rate, $\mdot \gtrsim 10^{-5}~\msunyr$,
which is ruled out by the weak H~I emission and lack of 
high ionization lines, such as He~II $\lambda$4686, on 
the optical spectrum (see Kenyon \& Webbink 1984).  
A `reprocessing' disk -- which absorbs and reradiates light 
from the primary -- can have a spectrum similar to a viscous 
accretion disk, because the temperature distributions
are similar (Kenyon \& Hartmann 1987).  Tests of several purely 
reprocessing disk models surrounding B-type stars indicate,
however, that the half-light radius of the disk is $\sim$ 
10 \rsun, instead of the $\sim$ 20 \rsun~derived from the 
optical eclipses.  We thus consider models where accretion 
and reprocessing contribute comparable amounts to the 
observed spectrum.  

The light solid line in Figure 8 plots the spectrum from 
an accretion disk that satisfies several observational
constraints.  The disk surrounds a B3 main sequence star.  
A large inclination, $i \ge$ 85\degree, allows the central 
star to hide behind the outer parts of the disk.  The disk 
may also shield the K0 star from the radiation field of the 
B star.  An accretion rate of a few $\times ~ 10^{-6} ~ \msunyr$ 
provides roughly half of the observed continuum radiation.  The model
predicts a disk temperature $\sim 10^4$ K at a disk radius
$\sim$ 20 \rsun~and a temperature $\sim$ 3000 K at $\sim$
80--90 \rsun.  These results are consistent with eclipse
data which require a small hot primary at short wavelengths
and a larger, cooler primary at long wavelengths (\S3.3).
Cool material in the outer disk may also produce the extra 
Mg~I absorption implied by the apparent rise in the Mg~I 
index at secondary minimum (Fig. 5.).  The uncertainties in 
this model are modest if we require the primary to lie on 
the main sequence.  Stars much cooler than $\sim$ 15,000 K 
(B6 V) produce disks that are too cold to match the optical 
spectrum.  Stars much hotter than $\sim$ 30,000 K (B0) should 
produce substantial He~I emission (from ionization of 
the disk) that is not observed.  

This model has one main failure.  It is difficult for a B3 star
to produce enough photons to ionize H to the observed level.  
We estimate an extinction-corrected H$\beta$ luminosity of 
$0.2~\lsun$, which requires $\sim 4 \times 10^{45}$ H-ionizing 
photons s$^{-1}$ if the B star ionizes material in the disk.  
This limit is 1--2 orders of magnitude larger than expected 
from a B3 V star.  A boundary layer at the inner edge of the 
disk probably also produces too few ionizing photons, unless 
we have underestimated the accretion rate (see, for example,
\cite{ken91}).  Olson (1991) considered H$\alpha$ line 
formation within the disks of several Algol systems, and 
produced line profiles similar to those observed.  It may 
be possible to account for the line flux in a similar way,
but such a calculation is beyond the scope of this paper.

It is unlikely that any B-type star in BG Gem could be
underluminous compared to a normal main sequence star.
Main sequence stars accreting material from a disk are
at least as luminous as non-accreting main sequence stars
of the same mass (e.g., Kippenhahn \& Meyer-Hofmeister 1977).

The best alternative to this model is a black hole primary.
Although less likely than a B-type primary star, a black hole 
accreting at a small fraction of the Eddington limit might 
account for the observed radiation at 20 \rsun~(depending on 
the geometry).  To evaluate the plausibility of this model,
we considered a disk surrounding a 4.5 \msun~black hole.  The 
thick, inner disk produces a luminosity of $\sim 10^3 ~ \lsun$ 
if the accretion rate is 1\% of the Eddington limit.  We naively 
assumed that the inner disk emits X-rays isotropically; the 
outer disk absorbs and re-radiates this emission at the local 
blackbody temperature.  For a steady-state disk where 
$H_d \propto R^{-9/8}$, the disk temperature at 4--20 \rsun~is 
$\sim$ 5000--8000 K.  The predicted luminosity of the outer disk 
in this model, $\sim 100 ~ \lsun$, is close to our estimated disk 
luminosity of 5--10 \lsun~if the inclination is $\sim$ 85\degree.

Our simple example suggests that a black hole accretion disk
can plausibly produce the disk emission observed in BG Gem for
accretion rates much smaller than the Eddington limit. The model 
optimistically assumes that the inner disk radiates efficiently,
$\sim$ 10\%, compared to advection-dominated accretion flows 
(e.g., \cite{esi97}) and that the outer disk radiates as a 
blackbody.  Despite this optimism, we note that Esin \etal (1997) 
derive optical luminosities, $\lambda F_{\lambda}$, of $\sim$
$10^{32}$--$10^{33}$ erg s$^{-1}$ for quiescent disks and
$10^{33}$--$10^{34}$ erg s$^{-1}$ for low state disks
in black hole binaries with inclinations of 60\degree--90\degree. 
The compact disks assumed in the Esin \etal models are more than 
factor of 100 smaller than the disk in BG Gem, so it may be
possible to achieve the observed optical flux from a larger 
disk surrounding a black hole in a wider binary system.

The black hole model has one main advantage and one main
disadvantage compared to models where the primary is a 
B-type main sequence star.  A black hole accreting at
1\% of the Eddington limit easily produces enough high
energy photons to satisfy our constraints on the H$\alpha$
and H$\beta$ fluxes unless the accretion efficiency is much
less than 1\% (Esin \etal 1997).  But, hiding an accreting 
black hole in an edge-on binary may be more difficult than 
hiding a B-type main sequence star.  The X-ray luminosity,
$\sim$ 1300 \lsun, needed to power the observed optical
luminosity in BG Gem is roughly 3 orders of magnitude larger 
than the upper limit derived from the ASM observations, $\sim$ 
2 \lsun.  The X-ray luminosities of quiescent black hole binary
systems are also low, $\lesssim~1~\lsun$ (e.g., \cite{mcc95}).
However, the region that emits X-rays in a black hole binary 
is very small, $\sim 10^{10}$ to $10^{11}$ cm, and easily 
occulted when the binary is viewed at high inclination. 
For example, Esin \etal (1997) show that the X-ray luminosity 
of the small disk in an edge-on black hole binary system 
is several orders of magnitude smaller than the same disk 
viewed pole-on.  We expect that the larger disk in BG Gem
can occult a larger fraction of the X-ray flux than an
edge-on disk in a smaller binary system.  Future calculations 
of black hole binaries should address this issue.

\section{DISCUSSION AND SUMMARY}

Our results indicate that BG Gem is a textbook example of
a very evolved, semi-detached binary system.  The system
consists of a low mass, lobe-filling K0 supergiant which
transfers material into a disk surrounding a massive primary
star.  The large ellipsoidal light variations of the K0
secondary indicate a large mass ratio, $q \approx$ 0.1,
and large orbital inclination, $i \gtrsim 80\degree$.  
The radial velocities of H$\alpha$ emission in the disk
support the large mass ratio.  We prefer $i \gtrsim 85\degree$ 
to allow the disk to occult the primary star, which is not
visible in the optical.  

The system is remarkably stable for an interacting binary.
There is little scatter in either the optical brightness 
(for 25 orbital cycles) or the emission line equivalent 
widths (for 3--4 orbital cycles).  Studies of archival
plate material would place useful constraints on long-term
variability and outburst frequency.

Our data are insufficient to choose between the two possible
alternatives for the primary star.  The optical spectra 
resemble spectra of several Algol systems -- e.g., 
UU Cnc (P = 96.7 days, \cite{eat91}) and 
RZ Oph (P = 262 days, \cite{bal78}; \cite{kne86}; \cite{zol91}) -- 
as well as quiescent spectra of several black hole binaries -- e.g., 
V616 Mon (P = 0.32 days, \cite{mcc95} and references therein) 
and V404 Cyg (P = 6.5 days, \cite{cas93} and references therein).
None of the long-period Algol binaries have as extreme mass
ratios as in BG Gem (see \cite{bat89}), although RZ Oph 
has $q \approx 0.2$.   The structure of the binary in RZ Oph 
has several other features in common with BG Gem, including 
a lobe-filling K-type star, an invisible primary, and an
extended accretion disk that occults the secondary star at
$\phi$ = 0.5 (see Olson 1987).  However, most of the energy 
from the disk in RZ Oph is produced by accretion, because the 
disk in RZ Oph is brighter and less stable than the disk 
in BG Gem.   Producing a large H$\alpha$ flux from the 
intrinsically less luminous disk in BG Gem is also a challenge,
as noted above.

BG Gem has many features in common with quiescent black hole
binaries, despite the 1--2 order of magnitude difference in
orbital periods.  Many black hole binaries have ellipsoidal 
light curves and extreme mass ratios similar to those in
BG Gem (e.g., \cite{bee97}; \cite{sha96}; \cite{sha94}).
Quiescent black hole binaries often have modest emission
line spectra, with strong, double-peaked H~I lines and
weaker, double-peaked He~I lines (e.g., \cite{cas93}; 
\cite{fil95}; \cite{mcc95}; \cite{oro98}).  Higher ionization 
emission lines, such as He II $\lambda$4686, appear only in 
outburst, when the accretion rate through the disk increases 
dramatically.  Quiescent black hole binaries often escape 
detection as very intense X-ray sources until they undergo 
major eruptions, and they often remain in quiescence for 
decades (e.g., \cite{tan95}; \cite{ued98}).  All of the 
currently known black hole binary systems have small orbital 
inclinations $i \lesssim 70\degree$ (see \cite{che97}; \cite{oro98}; 
\cite{fil99}).  The $i \gtrsim 85\degree$ derived here for 
BG Gem may allow the disk to occult even an intense X-ray
source (see \cite{so99a}, 1999b, and references therein
for typical outburst fluxes). Despite the lack of X-rays, 
BG Gem may still contain a black hole primary surrounded 
by a quiescent accretion disk.

If confirmed, BG Gem would be the black hole binary with
the longest known orbital period.  Other black hole binaries
have periods of a few days or less.  Any black hole binary
is a challenge to modern theories of the evolution of
binary systems (e.g., Kalogera 1999).  BG Gem would provide
a severe test of these theories.  BG Gem would also be the
only eclipsing system known among black hole binaries.
The black hole mass is unambiguous in an eclipsing binary.
Our limit for the primary mass is close to the maximum 
possible neutron star mass (see, for example, Burrows 1998).  
BG Gem thus may place new constraints on the minimum black 
hole mass.

With these considerations in mind, it is clear that 
additional observations are needed to understand the
nature of the primary in BG Gem.  Ultraviolet spectra
at 1000--3000 \AA~would probe the continuum and emission
lines produced in the inner portions of the accretion disk 
and might reveal stellar absorption lines if the primary is
a B-type star.  Better limits on the X-ray flux from
instruments on board {\it Chandra } and the {\it X-ray Multi-Mirror}
would also provide better constraints on the nature of the
primary.  Detection of an X-ray source that is eclipsed at
primary minima would favor a black hole primary, because
B-type main sequence stars are weak X-ray sources.  

Whatever the nature of the primary, future studies of the
secondary star and the accretion disk surrounding the primary 
should yield a better understanding of accretion in wide binary 
systems.  Eclipses of the blue continuum and the emission lines 
provide useful information about disk structure, through maximum 
entropy reconstruction among other techniques.  Higher resolution 
observations of Mg~I and other absorption features probe conditions 
in the secondary star as well as the cool, outer portions of the 
disk. Near-infrared observations covering the orbit would improve
mass estimates for the secondary star derived from ellipsoidal 
variations and might reveal emission from an optically thick 
``bright spot,'' where material lost by the K0 supergiant 
impacts the disk.  Emission from an optically thin spot might 
be visible in H$\alpha$ or H$\beta$.  Finally, high resolution 
observations might yield a radial velocity curve for the primary 
and improved estimates for the mass ratio.  The small distance 
and reddening make BG Gem a prime target for these and other 
observations of a binary system at an interesting phase of its evolution.

\vskip 6ex

We thank P. Berlind, M. Calkins, and other queue observers 
at the FLWO 60\inch~telescope for acquiring the FAST spectra 
used in this project. Susan Tokarz made the preliminary 
reductions of the FAST spectra.  We are grateful to Dan Fabricant 
and his excellent group for designing and building a wonderful 
optical spectrograph.  Comments from G. Sobczak, J. McClintock, 
R. Webbink, and an anonymous referee improved our presentation.
R. Remillard graciously derived the ASM upper limit.
We also thank the Wellesley College undergraduates who obtained the 
images of BG Gem for photometry.  We acknowledge summer support for 
A. B. (NSF grant AST-9417359)
and A. D. (W. M. Keck Foundation grant to the Keck Northeast
Astronomy Consortium). 
P. J. B. acknowledges support from NSF grant AST-9417359 and
the Brachman Hoffman Grants from Wellesley College.
This publication makes use of data products from the Two Micron All
Sky Survey, which is a joint project of the University of
Massachusetts and the Infrared Processing and Analysis Center, funded
by the National Aeronautics and Space Administration and the National
Science Foundation.
This research also used data through from the High Energy Astrophysics
Science Archive Research Center Online Service, provided by the NASA/Goddard
Space Flight Center.

\vfill
\eject
\singlespace

\begin{center}
\begin{deluxetable}{l c c c c c c c}
\singlespace
\scriptsize
\tablenum{1}
\tablecaption{Optical Photometry}
\tablehead{
\colhead{JD} & \colhead{Phase} &
\colhead{$\delta$V(BG)} & \colhead{$\delta$V(com)} &
\colhead{$\delta \rm R_C$(BG)} & \colhead{$\delta \rm R_C$(com)} &
\colhead{$\delta \rm I_C$(BG)} & \colhead{$\delta \rm I_C$(com)}}
\startdata
48694.5086 & 0.205 & $+$0.187 &  1.154 & $+$0.023 &  1.125 & $-$0.161 &  
1.122 \\
48698.5333 & 0.249 & $+$0.151 &  1.147 & $+$0.059 &  1.115 & $-$0.172 &  
1.122 \\
48984.5759 & 0.370 & $+$0.354 &  1.133 & $+$0.167 &  1.135 & $-$0.037 &  
1.118 \\
48990.7176 & 0.437 & $+$0.492 &  1.158 & $+$0.274 &  1.154 & $+$0.036 &  
1.141 \\
49006.5533 & 0.610 & $+$0.373 &  1.165 & $+$0.200 &  1.121 & $-$0.012 &  
1.111 \\
49044.5032 & 0.024 & $+$0.835 &  1.161 & $+$0.635 &  1.123 & $+$0.435 &  
1.130 \\
49084.5480 & 0.461 & $+$0.566 &  1.145 & $+$0.317 &  1.116 & $+$0.070 &  
1.147 \\
49106.5242 & 0.701 & $+$0.195 &  1.136 & $+$0.008 &  1.142 & $-$0.180 &  
1.134 \\
49368.7090 & 0.562 & $+$0.477 &  1.169 & $+$0.277 &  1.142 & $+$0.081 &  
1.123 \\
49375.6981 & 0.638 & $+$0.313 &  1.151 & $+$0.110 &  1.132 & $-$0.069 &  
1.125 \\
49382.7523 & 0.715 & \nodata & \nodata & $-$0.007 &  1.185 & $-$0.157 &  
1.126 \\
49390.6297 & 0.801 & $+$0.193 &  1.157 & $+$0.025 &  1.141 & $-$0.132 &  
1.112 \\
49391.5136 & 0.811 & $+$0.190 &  1.137 & $+$0.002 &  1.143 & $-$0.171 &  
1.127 \\
49398.5720 & 0.888 & $+$0.373 &  1.160 & $+$0.203 &  1.143 & $+$0.027 &  
1.119 \\
49402.7024 & 0.933 & $+$0.513 &  1.149 & $+$0.371 &  1.111 & $+$0.152 &  
1.131 \\
49406.6050 & 0.975 & $+$0.784 &  1.156 & $+$0.603 &  1.173 & $+$0.411 &  
1.125 \\
49409.4841 & 0.007 & $+$0.879 &  1.162 & $+$0.637 &  1.143 & $+$0.430 &  
1.136 \\
49410.6664 & 0.020 & $+$0.899 &  1.146 & $+$0.649 &  1.139 & $+$0.484 &  
1.119 \\
49411.5224 & 0.029 & $+$0.838 &  1.156 & $+$0.625 &  1.139 & $+$0.396 &  
1.155 \\
49412.5232 & 0.040 & $+$0.810 &  1.154 & $+$0.588 &  1.138 & $+$0.348 &  
1.143 \\
49413.5668 & 0.051 & $+$0.702 &  1.150 & $+$0.462 &  1.122 & $+$0.302 &  
1.118 \\
49416.5775 & 0.084 & $+$0.533 &  1.168 & $+$0.326 &  1.134 & $+$0.116 &  
1.118 \\
49423.6071 & 0.161 & $+$0.283 &  1.147 & $+$0.110 &  1.133 & $-$0.055 &  
1.123 \\
49435.6265 & 0.292 & \nodata & \nodata & $+$0.034 &  1.164 & $-$0.115 &  
1.120 \\
49438.5772 & 0.324 & $+$0.271 &  1.138 & $+$0.048 &  1.155 & $-$0.107 &  
1.118 \\
49442.5804 & 0.368 & $+$0.335 &  1.167 & $+$0.166 &  1.150 & $+$0.009 &  
1.116 \\
49443.6201 & 0.379 & $+$0.377 &  1.156 & $+$0.170 &  1.134 & $+$0.035 &  
1.103 \\
49444.5683 & 0.390 & $+$0.400 &  1.158 & $+$0.213 &  1.147 & $+$0.041 &  
1.132 \\
49447.5587 & 0.422 & $+$0.438 &  1.151 & $+$0.253 &  1.113 & $+$0.047 &  
1.132 \\
49461.5167 & 0.575 & $+$0.408 &  1.171 & $+$0.230 &  1.136 & $+$0.045 &  
1.131 \\
49744.6085 & 0.664 & $+$0.229 &  1.155 & $+$0.049 &  1.134 & $-$0.099 &  
1.114 \\
49746.6087 & 0.685 & $+$0.182 &  1.169 & $+$0.028 &  1.143 & $-$0.127 &  
1.116 \\
49747.6180 & 0.697 & $+$0.171 &  1.176 & $-$0.021 &  1.147 & $-$0.164 &  
1.120 \\
49748.5698 & 0.707 & $+$0.174 &  1.161 & $-$0.003 &  1.143 & $-$0.153 &  
1.116 \\
49751.6107 & 0.740 & $+$0.076 &  1.166 & $-$0.088 &  1.147 & $-$0.202 &  
1.102 \\
49752.5963 & 0.751 & $+$0.115 &  1.145 & \nodata & \nodata & $-$0.197 &  
1.114 \\
49755.5755 & 0.783 & $+$0.134 &  1.163 & $-$0.021 &  1.135 & $-$0.186 &  
1.111 \\
49757.5714 & 0.805 & $+$0.165 &  1.158 & $+$0.007 &  1.126 & $-$0.151 &  
1.110 \\
49761.5820 & 0.849 & $+$0.235 &  1.174 & $+$0.061 &  1.144 & $-$0.082 &  
1.122 \\
49762.5920 & 0.860 & $+$0.251 &  1.165 & $+$0.080 &  1.126 & $-$0.056 &  
1.112 \\
49763.5168 & 0.870 & $+$0.255 &  1.169 & $+$0.089 &  1.143 & $-$0.062 &  
1.121 \\
49766.5277 & 0.903 & $+$0.381 &  1.150 & $+$0.213 &  1.124 & $+$0.040 &  
1.125 \\
49771.5679 & 0.958 & $+$0.703 &  1.170 & $+$0.504 &  1.159 & $+$0.338 &  
1.118 \\
49780.6224 & 0.057 & $+$0.621 &  1.170 & $+$0.410 &  1.160 & $+$0.246 &  
1.111 \\
49781.6495 & 0.068 & $+$0.581 &  1.150 & $+$0.347 &  1.136 & $+$0.160 &  
1.124 \\
49804.5517 & 0.318 & $+$0.256 &  1.159 & $+$0.053 &  1.142 & $-$0.112 &  
1.125 \\
49810.5121 & 0.383 & $+$0.360 &  1.171 & $+$0.144 &  1.154 & $-$0.013 &  
1.117 \\
49815.5164 & 0.437 & $+$0.455 &  1.169 & $+$0.223 &  1.149 & $+$0.056 &  
1.109 \\
\enddata
\end{deluxetable}
\end{center}
\clearpage

\begin{center}
\begin{deluxetable}{l c c c c c c c}
\singlespace
\scriptsize
\tablenum{1}
\tablecaption{continued}
\tablehead{
\colhead{JD} & \colhead{Phase} &
\colhead{$\delta$V(BG)} & \colhead{$\delta$V(com)} &
\colhead{$\delta \rm R_C$(BG)} & \colhead{$\delta \rm R_C$(com)} &
\colhead{$\delta \rm I_C$(BG)} & \colhead{$\delta \rm I_C$(com)}}
\startdata
49831.5523 & 0.612 & $+$0.381 &  1.162 & $+$0.158 &  1.144 & $+$0.001 &  
1.117 \\
50114.6766 & 0.702 & $+$0.141 &  1.155 & $-$0.018 &  1.152 & $-$0.162 &  
1.119 \\
50118.6449 & 0.745 & $+$0.111 &  1.172 & $-$0.045 &  1.154 & $-$0.199 &  
1.131 \\
50120.6499 & 0.767 & $+$0.133 &  1.157 & $-$0.023 &  1.142 & \nodata & 
\nodata \\
50131.5678 & 0.886 & $+$0.310 &  1.162 & $+$0.145 &  1.141 & $-$0.022 &  
1.120 \\
50132.5850 & 0.897 & $+$0.342 &  1.165 & $+$0.186 &  1.140 & $+$0.016 &  
1.127 \\
50139.6648 & 0.974 & $+$0.702 &  1.172 & $+$0.523 &  1.148 & $+$0.353 &  
1.126 \\
50140.5501 & 0.984 & $+$0.736 &  1.157 & $+$0.548 &  1.140 & $+$0.383 &  
1.116 \\
50143.5953 & 0.017 & $+$0.777 &  1.171 & $+$0.587 &  1.149 & $+$0.403 &  
1.137 \\
50146.6049 & 0.050 & $+$0.646 &  1.161 & $+$0.456 &  1.140 & $+$0.266 &  
1.124 \\
50147.5501 & 0.060 & $+$0.602 &  1.174 & $+$0.394 &  1.139 & $+$0.207 &  
1.121 \\
50153.5759 & 0.126 & $+$0.317 &  1.161 & $+$0.136 &  1.139 & $-$0.046 &  
1.116 \\
50160.5511 & 0.202 & $+$0.128 &  1.179 & $-$0.034 &  1.156 & $-$0.198 &  
1.132 \\
50161.5227 & 0.213 & $+$0.142 &  1.170 & $-$0.025 &  1.157 & $-$0.186 &  
1.126 \\
50462.6408 & 0.499 & $+$0.633 &  1.166 & $+$0.392 &  1.153 & $+$0.185 &  
1.121 \\
50463.6463 & 0.509 & $+$0.620 &  1.167 & $+$0.398 &  1.142 & $+$0.178 &  
1.125 \\
50472.6658 & 0.608 & $+$0.369 &  1.155 & $+$0.160 &  1.140 & $-$0.015 &  
1.109 \\
50489.5630 & 0.792 & $+$0.137 &  1.147 & $-$0.013 &  1.138 & $-$0.175 &  
1.121 \\
50490.6027 & 0.804 & $+$0.136 &  1.161 & \nodata & \nodata & \nodata & 
\nodata \\
50504.5811 & 0.956 & $+$0.679 &  1.167 & $+$0.494 &  1.120 & $+$0.288 &  
1.162 \\
50523.5677 & 0.163 & $+$0.241 &  1.124 & $+$0.107 &  1.101 & $-$0.117 &  
1.116 \\
50782.8269 & 0.992 & $+$0.792 &  1.158 & $+$0.603 &  1.138 & $+$0.431 &  
1.122 \\
50828.6566 & 0.492 & $+$0.644 &  1.153 & $+$0.398 &  1.134 & $+$0.170 &  
1.125 \\
50836.5599 & 0.579 & $+$0.374 &  1.159 & $+$0.177 &  1.143 & $+$0.015 &  
1.132 \\
50839.5264 & 0.611 & $+$0.268 &  1.144 & $+$0.140 &  1.118 & $-$0.045 &  
1.107 \\
50840.5498 & 0.622 & $+$0.329 &  1.164 & $+$0.125 &  1.124 & $-$0.069 &  
1.122 \\
50858.6164 & 0.819 & $+$0.163 &  1.164 & $+$0.022 &  1.154 & $-$0.163 &  
1.131 \\
50860.6880 & 0.842 & $+$0.224 &  1.158 & $+$0.064 &  1.140 & $-$0.085 &  
1.099 \\
50883.6207 & 0.092 & $+$0.407 &  1.144 & $+$0.218 &  1.131 & $+$0.052 &  
1.132 \\
50897.5577 & 0.244 & $+$0.156 &  1.163 & $-$0.008 &  1.131 & $-$0.165 &  
1.124 \\
50898.5383 & 0.255 & \nodata & \nodata & $-$0.016 &  1.134 & $-$0.168 &  
1.119 \\
50903.5061 & 0.309 & $+$0.173 &  1.151 & $+$0.005 &  1.149 & $-$0.149 &  
1.134 \\
50914.5891 & 0.430 & $+$0.432 &  1.161 & $+$0.221 &  1.147 & $+$0.036 &  
1.148 \\
50915.5573 & 0.441 & $+$0.453 &  1.168 & $+$0.247 &  1.145 & $+$0.057 &  
1.126 \\
50931.5498 & 0.615 & \nodata & \nodata & $+$0.077 &  1.175 & $-$0.077 &  
1.142 \\
50932.5137 & 0.626 & $+$0.251 &  1.168 & $+$0.068 &  1.158 & $-$0.095 &  
1.122 \\
51210.6585 & 0.661 & $+$0.262 &  1.142 & $+$0.059 &  1.137 & $-$0.116 &  
1.136 \\
51219.5829 & 0.758 & $+$0.145 &  1.159 & $-$0.017 &  1.143 & $-$0.179 &  
1.121 \\
51225.5590 & 0.823 & $+$0.192 &  1.157 & $+$0.024 &  1.147 & $-$0.136 &  
1.123 \\
51226.6063 & 0.835 & $+$0.213 &  1.166 & $+$0.050 &  1.147 & $-$0.097 &  
1.121 \\
51233.5798 & 0.911 & $+$0.426 &  1.167 & $+$0.265 &  1.133 & $+$0.089 &  
1.122 \\
51240.5426 & 0.987 & $+$0.844 &  1.151 & $+$0.626 &  1.140 & $+$0.444 &  
1.107 \\
51248.5614 & 0.074 & $+$0.487 &  1.135 & $+$0.245 &  1.145 & $+$0.019 &  
1.111 \\
51255.5686 & 0.151 & $+$0.381 &  1.140 & $+$0.195 &  1.154 & $-$0.028 &  
1.128 \\
51263.4974 & 0.237 & $+$0.206 &  1.165 & $+$0.001 &  1.144 & $-$0.174 &  
1.128 \\
51269.5027 & 0.303 & $+$0.250 &  1.150 & $+$0.064 &  1.129 & $-$0.139 &  
1.125 \\
51273.5037 & 0.346 & $+$0.309 &  1.158 & $+$0.113 &  1.148 & $-$0.059 &  
1.131 \\
51277.6396 & 0.392 & $+$0.309 &  1.192 & $+$0.107 &  1.171 & $-$0.033 &  
1.133 \\
\enddata
\end{deluxetable}
\end{center}

\begin{center}
\begin{deluxetable}{l c c c c c}
\tablewidth{15cm}
\tablenum{2}
\tablecaption{Definitions of Spectral Indices}
\tablehead{
\colhead{Feature} & \colhead{Name} & 
\colhead{~~~~$\lambda$~~~~} & \colhead{$\lambda_b$} & 
\colhead{$\lambda_r$} & \colhead{$\delta \lambda$}}
\startdata
He~I      & I$_{\rm He}$       & 6678 & 6600 & 6750 & 30 \\
H$\alpha$ & I$_{\rm H \alpha}$ & 6563 & 6525 & 6600 & 30 \\
Ba I      & I$_{\rm Ba}$       & 6495 & 6475 & 6525 & 25 \\
Na I      & I$_{\rm Na}$       & 5893 & 5825 & 5965 & 30 \\
Mg I      & I$_{\rm Mg}$       & 5175 & 5050 & 5300 & 30 \\
H$\beta$  & I$_{\rm H\beta}$ & 4861 & 4825 & 4900 & 30 \\
Fe I      & I$_{\rm Fe}$       & 4392 & 4364 & 4448 & 35 \\
\\
7025 cont & $m_{7025}$ & 7025 & \nodata & \nodata & 30 \\
6370 cont & $m_{6370}$ & 6370 & \nodata & \nodata & 30 \\
5550 cont & $m_{5550}$ & 5550 & \nodata & \nodata & 30 \\
4400 cont & $m_{4400}$ & 4400 & \nodata & \nodata & 30 \\
4050 cont & $m_{4050}$ & 4050 & \nodata & \nodata & 30 \\
\enddata
\end{deluxetable}
\end{center}

\begin{center}
\begin{deluxetable}{l c c c c c c c}
\singlespace
\scriptsize
\tablewidth{15cm}
\tablenum{3}
\tablecaption{Absorption and Emission Line Indices }
\tablehead{
\colhead{JD} & \colhead{Phase} & 
\colhead{$I_{\rm Fe~I}$} & \colhead{$I_{\rm Mg~I}$} &
\colhead{$I_{\rm Na~I}$} & \colhead{$I_{\rm Ba~I}$} &
\colhead{EW$(H\beta)$}   & \colhead{EW$(H\alpha)$} }
\startdata
51070.0128 & 0.626 & 0.083 & 0.144 & 0.176 & 0.045 &  $-$8.5 & $-$27.7 \\
51070.0150 & 0.626 & 0.103 & 0.150 & 0.182 & 0.041 &  $-$9.1 & $-$28.3 \\
51086.0142 & 0.801 & 0.107 & 0.154 & 0.179 & 0.040 &  $-$7.9 & $-$25.8 \\
51101.0046 & 0.964 & 0.130 & 0.179 & 0.177 & 0.052 &  $-$5.8 & $-$20.1 \\
51102.9545 & 0.985 & 0.148 & 0.205 & 0.177 & 0.047 &  $-$2.6 & $-$16.4 \\
51109.9734 & 0.062 & 0.115 & 0.158 & 0.174 & 0.040 &  $-$9.5 & $-$27.4 \\
51115.9804 & 0.128 & 0.092 & 0.151 & 0.174 & 0.045 &  $-$8.6 & $-$28.1 \\
51129.9315 & 0.280 & 0.107 & 0.141 & 0.149 & 0.035 &  $-$6.6 & $-$23.8 \\
51132.9985 & 0.313 & 0.124 & 0.143 & 0.164 & 0.037 &  $-$7.0 & $-$25.8 \\
51135.9320 & 0.345 & 0.110 & 0.147 & 0.155 & 0.038 &  $-$7.0 & $-$26.4 \\
51138.9114 & 0.378 & 0.115 & 0.151 & 0.154 & 0.041 &  $-$7.6 & $-$27.9 \\
51141.9962 & 0.411 & 0.112 & 0.151 & 0.169 & 0.047 &  $-$9.5 & $-$29.9 \\
51144.8588 & 0.443 & 0.104 & 0.186 & 0.184 & 0.040 &$-$10.5 & $-$35.5 \\
51147.9806 & 0.477 & 0.111 & 0.170 & 0.197 & 0.042 &$-$13.0 & $-$40.7 \\
51157.7598 & 0.583 & 0.097 & 0.158 & 0.177 & 0.049 &$-$10.7 & $-$32.0 \\
51160.8444 & 0.617 & 0.106 & 0.144 & 0.160 & 0.042 &  $-$9.2 & $-$29.1 \\
51163.9600 & 0.651 & 0.096 & 0.139 & 0.156 & 0.029 &  $-$8.7 & $-$28.2 \\
51166.8621 & 0.683 & 0.102 & 0.154 & 0.159 & 0.048 &  $-$7.8 & $-$27.0 \\
51169.8996 & 0.716 & 0.103 & 0.150 & 0.160 & 0.045 &  $-$7.0 & $-$25.2 \\
51172.8716 & 0.748 & 0.109 & 0.151 & 0.163 & 0.047 &  $-$7.6 & $-$26.0 \\
51175.9043 & 0.781 & 0.092 & 0.143 & 0.154 & 0.043 &  $-$8.3 & $-$27.9 \\
51187.7566 & 0.911 & 0.094 & 0.164 & 0.157 & 0.048 &$-$10.7 & $-$29.4 \\
51188.7424 & 0.922 & 0.121 & 0.161 & 0.165 & 0.044 &  $-$9.4 & $-$27.9 \\
51191.7695 & 0.955 & 0.140 & 0.176 & 0.160 & 0.051 &  $-$6.4 & $-$21.1 \\
51194.8215 & 0.988 & 0.135 & 0.201 & 0.183 & 0.060 &  $-$2.0 & $-$15.3 \\
51196.7938 & 0.009 & 0.143 & 0.187 & 0.187 & 0.049 &  $-$2.9 & $-$15.9 \\
51198.7117 & 0.030 & 0.129 & 0.176 & 0.174 & 0.052 &  $-$5.1 & $-$18.9 \\
51201.7889 & 0.064 & 0.110 & 0.160 & 0.151 & 0.046 &  $-$9.8 & $-$27.6 \\
51216.6831 & 0.226 & 0.108 & 0.129 & 0.148 & 0.049 &  $-$7.3 & $-$24.6 \\
51218.6966 & 0.248 & 0.108 & 0.125 & 0.153 & 0.046 &  $-$6.9 & $-$24.3 \\
51223.5895 & 0.302 & 0.100 & 0.141 & 0.157 & 0.043 &  $-$7.1 & $-$25.8 \\
51224.6268 & 0.313 & 0.106 & 0.129 & 0.151 & 0.042 &  $-$7.0 & $-$25.7 \\
51226.7300 & 0.336 & 0.105 & 0.122 & 0.145 & 0.035 &  $-$6.8 & $-$29.7 \\
51228.6119 & 0.357 & 0.083 & 0.144 & 0.127 & 0.043 &  $-$7.2 & $-$28.1 \\
51230.7042 & 0.379 & 0.114 & 0.130 & 0.155 & 0.041 &  $-$8.3 & $-$29.1 \\
51232.7299 & 0.402 & 0.101 & 0.131 & 0.154 & 0.048 &  $-$8.8 & $-$30.6 \\
51247.7040 & 0.565 & 0.100 & 0.160 & 0.170 & 0.044 & $-$12.8 & $-$36.0 \\
51249.5973 & 0.586 & 0.100 & 0.150 & 0.160 & 0.043 & $-$11.0 & $-$33.2 \\
51257.6133 & 0.673 & 0.105 & 0.134 & 0.156 & 0.046 &  $-$8.6 & $-$27.5 \\
51259.6068 & 0.695 & 0.102 & 0.153 & 0.159 & 0.043 &  $-$8.0 & $-$27.2 \\
51278.6080 & 0.902 & 0.122 & 0.143 & 0.146 & 0.051 &  $-$9.1 & $-$28.2 \\
51280.6209 & 0.924 & 0.116 & 0.151 & 0.163 & 0.052 &  $-$8.7 & $-$25.7 \\
51284.6181 & 0.968 & 0.139 & 0.169 & 0.168 & 0.040 &  $-$3.3 & $-$18.1 \\
51287.6168 & 0.000 & 0.143 & 0.198 & 0.176 & 0.055 &  $-$1.7 & $-$14.0 \\
51289.6136 & 0.022 & 0.126 & 0.188 & 0.187 & 0.049 &  $-$4.1 & $-$17.6 \\
51291.6213 & 0.044 & 0.122 & 0.173 & 0.173 & 0.049 &  $-$8.3 & $-$22.8 \\
\enddata
\end{deluxetable}
\end{center}

\begin{center}
\begin{deluxetable}{l c c c c c c}
\singlespace
\tablewidth{15cm}
\tablenum{4}
\tablecaption{Continuum Magnitudes}
\tablehead{
\colhead{JD} & \colhead{Phase} & 
\colhead{$m_{4050}$} & \colhead{$m_{4400}$} &
\colhead{$m_{5550}$} & \colhead{$m_{6370}$} &
\colhead{$m_{7025}$}} 
\startdata
51101.0046 & 0.964 & 15.19 & 14.40 & 13.33 & 13.15 & 13.07 \\
51102.9545 & 0.985 & 15.34 & 14.59 & 13.47 & 13.35 & 13.20 \\
51109.9734 & 0.062 & 14.40 & 14.05 & 13.05 & 12.87 & 12.82 \\
51115.9804 & 0.128 & 14.35 & 14.00 & 13.06 & 12.89 & 12.80 \\
51135.9320 & 0.345 & 14.31 & 13.99 & 13.06 & 12.89 & 12.87 \\
51141.9962 & 0.411 & 14.29 & 14.06 & 13.18 & 13.05 & 13.07 \\
51157.7598 & 0.583 & 14.57 & 14.17 & 13.13 & 12.94 & 12.90 \\
51160.8444 & 0.617 & 14.43 & 14.07 & 13.09 & 12.94 & 12.93 \\
51172.8716 & 0.748 & 14.10 & 13.75 & 12.83 & 12.69 & 12.65 \\
51175.9043 & 0.781 & 14.21 & 13.86 & 12.91 & 12.76 & 12.74 \\
51188.7424 & 0.922 & 14.44 & 14.05 & 13.15 & 13.08 & 13.07 \\
51191.7695 & 0.955 & 15.02 & 14.55 & 13.45 & 13.26 & 13.20 \\
51196.7938 & 0.009 & 15.33 & 14.65 & 13.48 & 13.29 & 13.13 \\
51216.6831 & 0.226 & 14.32 & 13.91 & 13.07 & 12.91 & 12.77 \\
51218.6966 & 0.248 & 14.23 & 13.80 & 13.06 & 12.92 & 12.84 \\
51223.5895 & 0.302 & 14.31 & 13.90 & 13.04 & 12.90 & 12.85 \\
51230.7042 & 0.379 & 14.22 & 13.92 & 12.97 & 12.84 & 12.81 \\
51247.7040 & 0.565 & 14.70 & 14.32 & 13.33 & 13.15 & 13.15 \\
51259.6068 & 0.695 & 14.26 & 13.86 & 12.91 & 12.81 & 12.72 \\
51278.6080 & 0.902 & 14.45 & 14.07 & 13.16 & 13.06 & 13.01 \\
51280.6209 & 0.924 & 14.41 & 14.04 & 13.10 & 13.00 & 13.01 \\
51287.6168 & 0.000 & 15.36 & 14.65 & 13.47 & 13.34 & 13.22 \\
51289.6136 & 0.922 & 14.44 & 14.05 & 13.05 & 12.98 & 12.95 \\
51291.6213 & 0.944 & 14.77 & 14.28 & 13.24 & 13.06 & 13.01 \\
\enddata
\end{deluxetable}
\end{center}

\begin{center}
\begin{deluxetable}{l c r r}
\singlespace
\footnotesize
\tablewidth{15cm}
\tablenum{5}
\tablecaption{Radial Velocities}
\tablehead{
\colhead{JD} & \colhead{Phase} &
\colhead{$v_{abs}$} & \colhead{$v_{\rm H \beta}$}}
\startdata
2451070.0150 & 0.626 &$-$19.0 & 41.8 \\
2451086.0142 & 0.801 &$-$36.9 & 41.2 \\
2451101.0046 & 0.964 &   5.4  & 212.0 \\
2451102.9545 & 0.985 & $-$11.1 & 171.5 \\
2451109.9734 & 0.062 &  56.9 & 49.3 \\
2451115.9804 & 0.128 &  77.4 & 41.6 \\
2451129.9315 & 0.280 &  98.5 & 56.7 \\
2451132.9985 & 0.313 &  80.0 & 21.9 \\
2451135.9320 & 0.345 &  82.1 & 24.2 \\
2451138.9114 & 0.378 &  77.1 & 17.5 \\
2451141.9962 & 0.411 &  80.2 & 32.1 \\
2451144.8588 & 0.443 &  56.8 &  5.4 \\
2451147.9806 & 0.477 &  24.5 & 32.7 \\
2451157.7598 & 0.583 & $-$23.2 & 18.0 \\
2451160.8444 & 0.617 & $-$29.6 & 48.2 \\
2451163.9600 & 0.651 & $-$30.6 & 33.9 \\
2451166.8621 & 0.683 & $-$48.6 & 35.2 \\
2451169.8996 & 0.716 & $-$31.2 & 55.1 \\
2451172.8716 & 0.748 & $-$36.6 & 57.3 \\
2451175.9043 & 0.781 & $-$35.0 & 25.7 \\
2451187.7566 & 0.911 &  $-$5.2 & 61.9 \\
2451188.7424 & 0.922 &   7.3 & 73.0 \\
2451191.7695 & 0.955 &  12.7 & 207.9 \\
2451194.8215 & 0.988 &  31.0 & 153.7 \\
2451196.7938 & 0.009 &  42.6 & $-$46.3 \\
2451198.7117 & 0.030 &  50.2 & $-$99.2 \\
2451201.7889 & 0.064 &  87.1 & 23.8 \\
2451216.6831 & 0.226 & 129.0 & 36.0 \\
2451218.6966 & 0.248 & 118.0 & 66.0 \\
2451223.5895 & 0.302 &  95.3 & 28.7 \\
2451224.6268 & 0.313 & 125.0 & 31.7 \\
2451226.7300 & 0.336 & 104.3 & 63.1 \\
2451228.6119 & 0.357 &  41.8 & 27.6 \\
2451230.7042 & 0.379 &  68.4 & 38.7 \\
2451232.7299 & 0.402 &  43.8 & 16.8 \\
2451247.7040 & 0.565 & $-$10.3 & 4.5 \\
2451249.5973 & 0.586 & $-$21.4 & 16.0 \\
2451257.6133 & 0.673 & $-$54.3 & 12.0 \\
2451259.6068 & 0.695 & $-$40.9 & 43.0 \\
2451278.6080 & 0.902 &  $-$3.2 & 74.9 \\
2451280.6209 & 0.924 &  10.6 & 125.5 \\
2451284.6181 & 0.968 &  18.3 & 178.2 \\
2451287.6168 & 0.000 &  15.3 & 12.8 \\
2451289.6136 & 0.022 &  33.4 & $-$71.5 \\
2451291.6213 & 0.044 &  55.2 & $-$46.5 \\
\enddata
\end{deluxetable}
\end{center}

\end{document}